\documentclass[12pt,1p,review]{elsarticle}

\usepackage{graphicx}
\usepackage{dcolumn}
\usepackage{bm}
\usepackage{amssymb}
\usepackage{comment}
\usepackage{subfigure}
\usepackage{tikz}
\usetikzlibrary{scopes}
\usetikzlibrary{calc}
\usepackage{amsthm}
\usepackage{amsopn}
\usepackage{amsmath}

\def\bra#1{\mathinner{\langle{#1}|}}
\def\ket#1{\mathinner{|{#1}\rangle}}

\newcommand{\ve}[1]{\boldsymbol{#1}}

\journal{Chemical Physics}

\begin{document}

\title{Crystallographic refinement of collective excitations using standing wave inelastic x-ray scattering}%

\begin{frontmatter}

\author{Yu Gan}
\author{Anshul Kogar}
\author{Peter Abbamonte}
\address{Department of Physics and Federick Seitz Materials Research Laboratory, University of Illinois, Urbana, IL, 61801}
\address{Advanced Photon Source, Argonne National Laboratory, Argonne, IL, 60439}

\begin{abstract}
We propose a method for realizing true, real-space imaging of charge dynamics in a periodic system, with angstrom spatial resolution and attosecond time resolution.  In this method, inelastic x-ray scattering (IXS) is carried out with a coherent, standing wave source, which provides the off-diagonal elements of the generalized dynamic structure factor, $S(\ve{q}_1,\ve{q}_2,\omega)$, allowing complete reconstruction of the inhomogeneous response function of the system, $\chi(\ve{x}_1,\ve{x}_2,t)$.  The quantity $\chi$ has the physical meaning of a propagator for charge, so allows one to observe$-$in real time$-$the disturbance in the electron density created by a point source placed at a specified location, $\ve{x_1}$ (on an atom vs. between atoms, for example).  This method may be thought of as a generalization of x-ray crystallography that allows refinement of the excited states of a periodic system, rather than just its ground state.

\end{abstract}

\begin{keyword}
\end{keyword}

\end{frontmatter}

\section{Introduction}
\label{sec:intro}
Electron dynamics underlie all fundamental phenomena in chemistry, biology and materials physics.  Recent advances in attosecond laser sources have  created widespread interest in studying such electronic processes in real time, particularly in cases where spatial information, e.g., about the detailed configuration of electron wave packets, can be inferred.\citep{krausz,atto1,atto2}

We recently proposed a different approach to attosecond imaging based on inelastic x-ray scattering (IXS).\citep{advmat}  Unlike more common, time-domain approaches that exploit state-of-the-art laser technology, in the IXS approach an x-ray photon modulates the electron density of the system, and the ensuing response is measured the momentum and frequency domain.  Spatial and temporal information is then achieved by solving the inverse scattering problem, which for typical experimental setups yields time resolutions in the range of few attoseconds.\citep{advmat}  This method, which we will refer to here as ``IXS imaging," should be considered a member of the same broad technique class as x-ray crystallography, which is routinely used to obtain real space images$-$for example of protein structures$-$with angstrom resolution.\cite{rupp}  The IXS approach to studying dynamics is somewhat more restrictive than laser-based approaches, in that it strictly allows one to observe the time-evolution of the electron density of the system in response to a charged, point source.\cite{advmat}  However, it provides explicit, real-space images, and contains no intrinsic limit on the achievable time resolution: attosecond or even zeptosecond resolution is achievable.  So far, IXS imaging has been used to study collective electron dynamics in liquid water,\citep{water,coridan} excitons in large-gap insulators,\citep{LiF} and to measure the effective fine structure constant of graphene.\citep{reed}

Up to now, however, there has been a limitation on the imaging aspect of this approach.  The objective in IXS imaging is to determine the charge response function (or``propagator"), $\chi(\ve{x}_1,\ve{x}_2,t)$, which physically represents the amplitude that a point disturbance in the electron density at location $\ve{x}_1$ will propagate to $\ve{x}_2$ after elapsed time $t$.\footnote{By reciprocity, the probability of propagating instead from $\boldsymbol{x}_2$ to $\boldsymbol{x}_1$ is the same.} In reciprocal space this quantity is a function of two momenta, $\chi(\ve{q}_1,\ve{q}_2,\omega)$, but conventional IXS probes only its diagonal components, $\chi(\ve{q},-\ve{q},\omega)$.\citep{advmat} The Fourier transform of the latter quantity, $\chi(\ve{x},t)$, is causal and quantitatively accurate in its time evolution.  However, it depends on only one spatial variable, and corresponds to the complete response averaged over all source locations.\citep{abbamonte:054302}  If the system of interest is homogeneous, for example a free electron metal such as aluminum, this spatial averaging results in no loss of information.  However, in an inhomogeneous system, important, local features may be averaged out.

To overcome this limitation one must measure the off-diagonal momentum components of $\chi(\ve{q}_1,\ve{q}_2,\omega)$, i.e., where $\ve{q}_1 \neq -\ve{q}_2$.  It was shown many years ago that this is possible, at least in principle, by using coherent standing waves.\citep{schulke1,golovchenko}.  In this approach, an x-ray standing wave is created by exciting a Bragg reflection in a crystal (or by using an external coherent source, such as a Bonse-Hart interferometer).  Within a coherence volume, the x-ray photon lies in a quantum superposition of two distinct momenta, $\ve{k}_1$ and $\ve{k}_2$.  One then places the IXS detector (typically a backscattering analyzer) at some scattering angle, which defines {\it two} momentum transfers, $\ve{q}_1$ and $\ve{q}_2$. Under these conditions, due to interference between the two scattering channels, the cross section includes terms that depend on $\chi(\ve{q}_1,\ve{q}_2,\omega)$.

Could this approach be used to solve the averaging problem in IXS imaging?  It is not obvious that it can.  In standing wave techniques one does not have complete freedom to choose the two momenta, $\ve{k}_1$  and $\ve{k}_2$.  Because the wave field is created by a Bragg reflection, the momenta are always related by a reciprocal lattice vector, i.e., $\ve{k}_1 - \ve{k}_2 = \ve{G}$.  Hence, one's access to reciprocal space is highly constrained, and it is unclear whether enough information is accessible to permit a full refinement of $\chi(\ve{x}_1,\ve{x}_2,t)$.  

In this article we examine this issue by analyzing a simple model of a single quantum particle in a periodic potential.  We find that whether enough reciprocal space is accessible depends on the dimensionality of the problem.  We show that the standing wave approach fails in both one and two dimensions, because the dimensionality of the accessible momentum space is lower than the dimensionality of the set of data needed to perform a refinement.  In three dimensions, however, the technique should in principle be viable: by performing IXS measurements under a sufficiently large set of standing wave conditions$-$each defined by a distinct reciprocal lattice vector$-$ it should be possible to reconstruct the complete, unaveraged response $\chi(\ve{x}_1,\ve{x}_2,t)$.  In this sense, this approach can be thought of as a new type of x-ray crystallography that allows refinement of the collective excitations of a periodic system, rather than just its ground state.

\section{Background}

We begin by reviewing spatially-averaged IXS imaging, as it has been implemented previously\citep{water,LiF,reed} (for a more thorough review, see ref. \cite{advmat}).  In IXS, a monochromatic beam of x-rays is impinged on a system, from which it scatters in all directions.  An energy-resolving detector is placed at some angle and measures the spontaneously Raman scattered photons.  This detector defines a transferred energy, $\omega$, and transferred momentum, $\ve{q}$, and the intensity measured is proportional to the  dynamic structure factor, $S(\ve{q},\omega)$, which is the Fourier transform of the density-density correlation function\citep{sinha,advmat}. 

While $S$ contains a great deal of information about excited states, it is of rather limited interest from the point of view of dynamics.  As an autocorrelation function, the dynamical information contained in $S$ is indirect, reflecting only the general length and time scales over which excitations take place.  Fortunately, $S(\ve{q},\omega)$ is related to the response function, $\chi(\ve{q}_1,\ve{q}_2,\omega)$, which reveals the real space and time dynamics, by the quantum mechanical version of the fluctuation-dissipation theorem,
\begin{equation}
\label{eq:f-d}
S(\ve{q},\omega)=-\frac{1}{\pi}\frac{1}{1-e^{-\beta\hbar\omega}}\operatorname{Im}\chi(\ve{q},-\ve{q},\omega).
\end{equation}
Hence, in principle, IXS provides access to the true, causal dynamics of the system.  

\begin{figure}\centering
\includegraphics[width=0.45\textwidth]{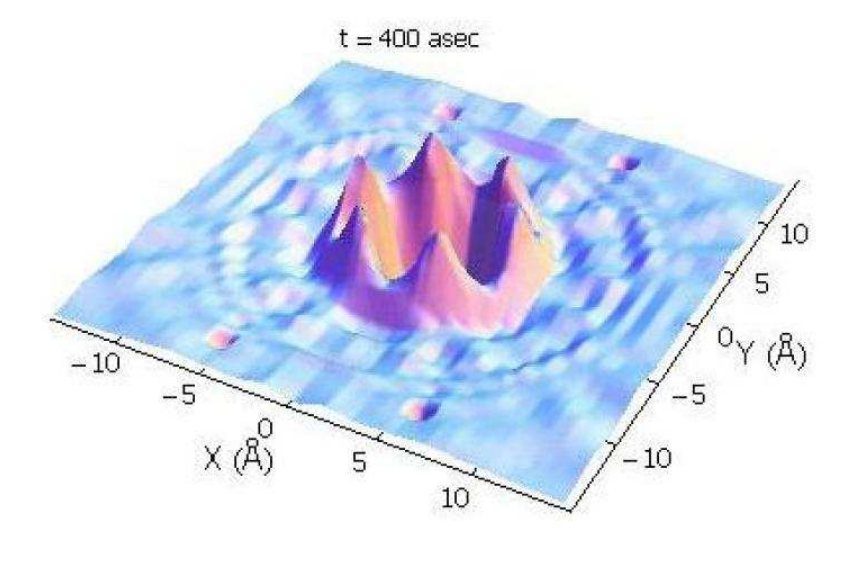}
\caption{$\chi(\ve{x},t)$ at $t=400$ as determined from a spatially-averaged IXS study of single crystal graphite, reproduced from \citep{reed}.}
\label{fig:graphite}
\end{figure}

Of course, $S$ does not supply the entire $\chi$ function, but only its imaginary part.  This restriction is the IXS rendition of the well-known phase problem in x-ray crystallography.\cite{rupp}  In the present case, the phase problem may be solved by recognizing that $\chi$ is a causal function, i.e., the real part can be obtained from the Kramers-Kronig (KK) relation
\begin{equation}
\label{eq:kramers}
\operatorname{Re}\chi(\ve{q},-\ve{q},\omega)=\frac{2}{\pi}\mathcal{P}\int_0^{\infty}{d\omega' \frac{\omega'\operatorname{Im}\chi(\ve{q},-\ve{q},\omega')}{(\omega')^2-\omega^2}},
\end{equation}
where $\mathcal{P}$ denotes the principal part. Inverse Fourier transforming gives a function, $\chi(\ve{x},t)$, that corresponds to the complete response, $\chi(\ve{x}_1,\ve{x}_2,t)$, averaged over all space with the difference $\ve{x}=\ve{x}_1-\ve{x}_2$ held fixed.\cite{abbamonte:054302} In a homogeneous system, $\chi(\ve{x},t)$ represents a complete parameterization of the response. Fig.~\ref{fig:graphite}, for example, shows an image of the electron density in graphite 400 attoseconds after the source.\cite{reed}

Note that Eq.~\ref{eq:kramers} implies a need for a significant quantity of experimental data.  The response $\chi(\ve{q},-\ve{q},\omega)$ is a strong function of the three-dimensional momentum, $\ve{q}$, and must be sampled with enough range and point density to define an inverse spatial Fourier transform.  Moreover, at every value of $\ve{q}$, a spectrum must be obtained over a sufficiently large range to carry out the KK transform, Eq.~\ref{eq:kramers}.  The number of unknowns can often be reduced by examining specific projections of reduced dimensionality \cite{LiF} or by exploiting crystal symmetry \cite{reed} (see section VI).  But, in general, this method requirs a large amount of data, and has only become possible because of the advent of high-brightness, third-generation synchrotron x-ray sources.

\section{IXS with standing waves}
\label{sec:cixs}

\begin{figure}\centering
\subfigure[][]{\includegraphics[width=0.5\textwidth]{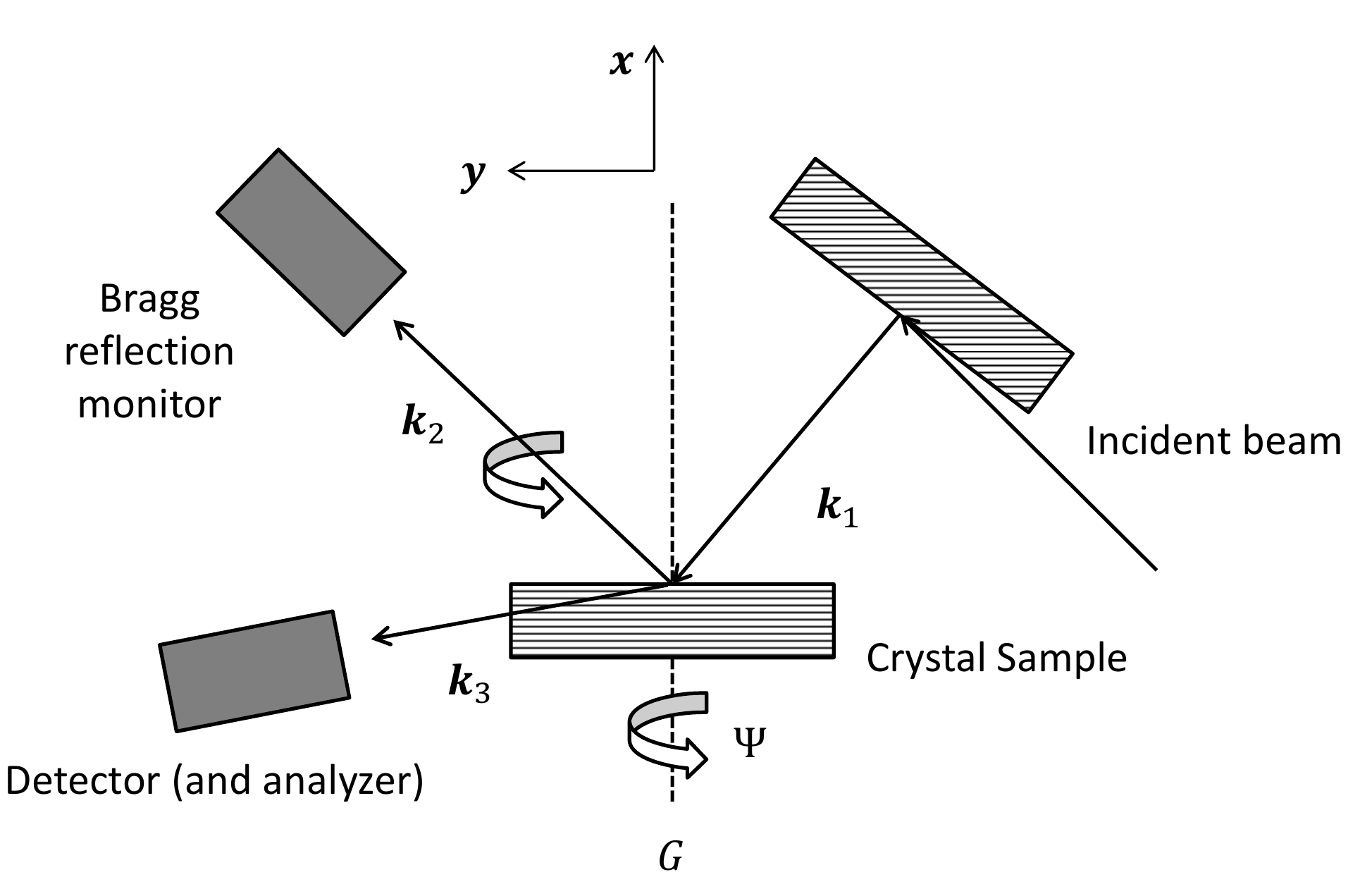}
\label{fig:exppic}}
\subfigure[][]{\includegraphics[width=0.4\textwidth]{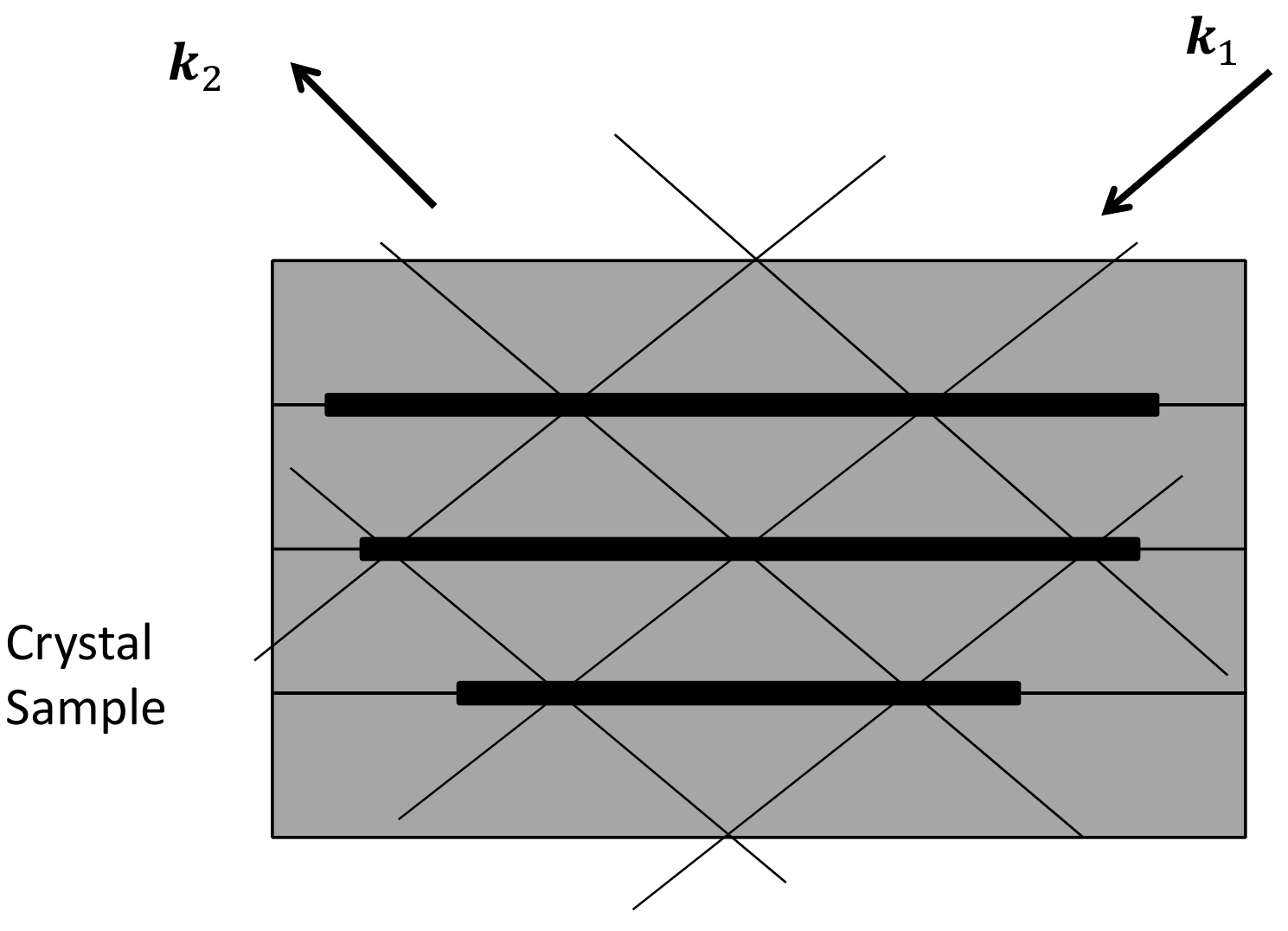}
\label{fig:standpic}}
\caption{\subref{fig:exppic} Layout of a standing wave IXS imaging experiment.  Two identical crystals of the system of interest are mounted nondispersively.  The first, which is asymmetrically cut, collimates the beam, and the second generates the standing wave.  
$\boldsymbol{k}_1$, $\boldsymbol{k}_2$, and $\ve{k}_3$ are the momenta of the incident, Bragg reflected, and inelastically scattered photons, respectively.  $\Psi$ describes rotations of the second crystal around $\ve{G}$.
\subref{fig:standpic} Standing wave-field formed by incident photon $\ve{k}_1$ and Bragg reflected photon $\ve{k}_2$.  Nodal planes of the standing wave-field are highlighted in blue.  The nodal planes are shown to be coincident with the lattice planes, but the phase of the standing wave-field, and hence the position of the nodes, is tunable.}
\label{fig:genexp}
\end{figure}

To obtain unaveraged images, i.e., that will give a faithful representation of the dynamics of systems that are inhomogeneous, the previous approach must be extended to determine the off-diagonal response, $\chi(\ve{q}_1,\ve{q}_2,\omega)$.  As discussed in Section 1, this might be accomplished by the use of an x-ray standing wave.  The most straight-forward way to produce a standing wave is by exciting a Bragg reflection.  Hence, we will from this point assume that the system of interest is available in the form of high quality single crystals, such as those commonly used in structural biology.\cite{rupp}

The proposed geometry for the experiment is illustrated in Fig.~\ref{fig:genexp}.  Two, identical crystals of the system of interest are mounted in the geometry shown, with the first cut asymmetrically with respect to a particular set of lattice planes, defined by reciprocal lattice vector, $\ve{G}$, and the second cut symmetrically.  As described by dynamical diffraction theory,\cite{als-nielsen} the first crystal compresses the angular divergence of the beam so that it is much less than the Darwin width, $\theta_W$, of the second, symmetric reflection.  Under these circumstances, the second crystal will generate a coherent standing wave, in which the photon lies in a quantum superposition of two momenta, described by the quantum state \citep{schulke1,golovchenko,schulke2}  
\begin{align}
\label{eq:initial}
\ket{i} & = (g_1 a^\dagger_{\ve{k}_1\alpha_1}+ g_2 a^\dagger_{\ve{k}_2\alpha_2}e^{i\gamma})\ket{m}.
\end{align}
In this expression $g_1$ and $g_2$ are the amplitudes of the incident and diffracted beams, respectively, and $\gamma$ is the phase shift between the two.  $a_{\ve{k}\alpha}$ annihilates a photon with momentum $\ve{k}$ and polarization state $\alpha$, and $\ket{m}$ is the initial many-body state of the electron system.  Note that, because the wave field is created by a Bragg reflection, $\ve{k}_1-\ve{k}_2=\ve{G}$.  Nevertheless, because Bragg scattering is an elastic process, $\omega(\ve{k}_1)=\omega(\ve{k}_2)$.  For notational simplicity, we take $\omega_1 \equiv \omega(\ve{k}_1)=\omega(\ve{k}_2)$ and $\omega_3 \equiv \omega(\ve{k}_3)$ (Fig. 2) .  The parameters $g_2$ and $\gamma$ may be controlled by fine adjustment of the angle of either crystal, and both quantities may be accurately calculated from dynamical diffraction theory.  For a thorough review of x-ray standing wave techniques, we refer the reader to Refs. \citep{als-nielsen,bedzyk}.

Once the standing wave is established, the initial state in the inelastic scattering process is that shown in Eq. \ref{eq:initial}.  The final state is an ordinary plane wave state, given by 
\begin{align}
\label{eq:final}
\ket{f} & = a^{\dagger}_{\ve{k}_3\alpha_3}\ket{n}.
\end{align}
where $\ve{k}_3$ is the wave vector of the inelastically scattered photon, and $\ket{n}$ is the final many body state to which the electron system was excited.

The scattering from $\ket{i}$ to $\ket{f}$ is mediated by the usual interaction between light and matter.  The Hamiltonian for a system of non-relativistic electrons interacting with the  electromagnetic field is \cite{sakurai}
\begin{align}
\hat H = {\hat H_0} + \frac{e}{2mc} \int{ \hat{\psi}^\dagger  \hat{\ve{A}}\cdot \ve{p} \, \hat{\psi} \, d\ve{x}}
+ \frac{e^2}{2mc^2}\int{ \hat{\rho} \,  \hat{\ve{A}}^2 \, d\ve{x}}
\end{align}


\noindent
where $\hat{H}_0$ is the many-body Hamiltonian of the electron system, $\hat{\psi}(\ve{x})$ is the electron field operator that annihilates an electron at position $\ve{x}$, and $\hat{\rho}(\ve{x})=\hat{\psi}^\dagger(\ve{x}) \hat{\psi}(\ve{x})$ is the electron density operator.  $\hat{\ve{A}}$ is the quantized vector potential,
\begin{align}
\hat{\ve{A}}(\ve{x}) & = \left(\frac{2\pi\hbar}{V}\right)^{\frac{1}{2}}\displaystyle\sum_{k\alpha}\frac{c}{\sqrt{\omega_k}}\left(\hat{\ve{\epsilon}}^*_{\ve{k}\alpha}
a^\dagger_{\ve{k}\alpha}e^{-i\ve{k}\cdot\ve{x}} +\hat{\ve{\epsilon}}_{\ve{k}\alpha}a_{\ve{k}\alpha}e^{i\ve{k}\cdot\ve{x}}\right),\label{eq:gauge}
\end{align}

\noindent
which either creates or annihilates a photon at location $\ve{x}$. 

To lowest order in perturbation theory, only the second interaction, $H_{\textrm{int}}=(e^2/2mc^2) \int \hat{\rho} \hat{A}^2 d\ve{x}$, leads to scattering.\cite{sinha}  The first order scattering amplitude is given by 
\begin{align}
\bra{f}H_{\textrm{int}}\ket{i}  = & \frac{2\pi\hbar e^2}{mV}\frac{g_1}{\sqrt{\omega_3\omega_1}} (\hat{\ve{\epsilon}}^*_{3}\cdot\hat{\ve{\epsilon}}_{1})
\bra{ n }\rho(-\ve{q}_1)\ket{m}  \nonumber \\
&  + \frac{2\pi\hbar e^2}{mV}\frac{g_2 e^{i\gamma}}{\sqrt{\omega_3\omega_2}} (\hat{\ve{\epsilon}}^*_{3}\cdot\hat{\ve{\epsilon}}_{2})
\bra{ n }\rho(-\ve{q}_2)\ket{m}.\label{eq:predd}
\end{align}
where we have defined the two momentum transfers $\ve{q}_1=\ve{k}_1-\ve{k}_3$ and $\ve{q}_2=\ve{k}_2-\ve{k}_3$.  The transition rate $\Gamma_{i\rightarrow f}$ at finite temperature $T$ is given by Fermi's golden rule
\begin{align}
\Gamma_{i\rightarrow f}  = & \frac{4\pi^2\hbar}{\omega_1\omega_3}\left(\frac{e^2}{mV}\right)^2\displaystyle\sum_{n,m} b_m  \left[g_1^2 |\hat{\ve{\epsilon}}^*_{3}\cdot\hat{\ve{\epsilon}}_{1}|^2|\bra{ n }\rho(\ve{q}_1)\ket{m}|^2 \right. \nonumber \\
& \left. + g_2^2 |\hat{\ve{\epsilon}}^*_{3}\cdot\hat{\ve{\epsilon}}_{2}|^2|\bra{ n }\rho(\ve{q}_2)\ket{m}|^2 \right. \nonumber \\
& \left. + g_1 g_2 e^{i\gamma} (\hat{\ve{\epsilon}}_{3}\cdot\hat{\ve{\epsilon}}^*_{1})(\hat{\ve{\epsilon}}^*_{3}\cdot\hat{\ve{\epsilon}}_{2})\bra{m}\rho(\ve{q}_1)\ket{n}\bra{ n }\rho(-\ve{q}_2)\ket{m} \right. \nonumber \\
& \left. + g_1 g_2 e^{-i\gamma} (\hat{\ve{\epsilon}}^*_{3}\cdot\hat{\ve{\epsilon}}_{1})(\hat{\ve{\epsilon}}_{3}\cdot\hat{\ve{\epsilon}}^*_{2})\bra{m}\rho(\ve{q}_2)\ket{n}\bra{ n }\rho(-\ve{q}_1)\ket{m}\right] \nonumber \\
& \times \delta(E-E_n+E_m),
\label{eq:fermi}
\end{align}
where we have used the fact that the electron density is real, i.e., $\hat{\rho}^\dagger(\ve{q})=\hat{\rho}(-\ve{q})$.  Here $b_m = e^{-\beta E_m}/Z$ is the Boltzmann factor and $E$ is the energy transferred to the sample by the photon.

The quantity that is relevant to the experiment is the doubly-differential scattering cross section\cite{sakurai}
\begin{equation}
\frac{\partial^2\sigma}{\partial\Omega \partial E}= \frac{1}{\Phi}\Gamma_{i\rightarrow f} \frac{\partial^2 N}{\partial \Omega\partial E},
\label{eq:ddcsbasic}
\end{equation}
where $\partial^2 N / \partial \Omega\partial E$ is the density of final states and $\Phi$ is the incident flux.  Because the final state contains a single photon with polarization $\alpha$, we have $\partial^2 N/\partial \Omega\partial E = V \omega_3^2/8\pi^3\hbar c^3$ and $\Phi=c/V$.  Substituting into Eq.~\ref{eq:ddcsbasic}, we have
\begin{align}
\frac{d^2\sigma}{dE' d\Omega}=& \left(\frac{e^2}{mc^2}\right)^2\frac{\omega_3}{\omega_1}\displaystyle\sum_{n,m} b_m \left[g_1^2 |\hat{\ve{\epsilon}}^*_{3}\cdot\hat{\ve{\epsilon}}_{1}|^2|\bra{ n }\rho(\ve{q}_1)\ket{m}|^2 \right. \nonumber \\
& \left. + g_2^2 |\hat{\ve{\epsilon}}^*_{3}\cdot\hat{\ve{\epsilon}}_{2}|^2|\bra{ n }\rho(\ve{q}_2)\ket{m}|^2 \right. \nonumber \\
& \left. + g_1 g_2 e^{i\gamma}(\hat{\ve{\epsilon}}_{3}\cdot\hat{\ve{\epsilon}}^*_{1})(\hat{\ve{\epsilon}}^*_{3}\cdot\hat{\ve{\epsilon}}_{2})\bra{m}\rho(\ve{q}_1)\ket{n}\bra{n}\rho(-\ve{q}_2)\ket{m} \right. \nonumber \\
& \left. + g_1 g_2 
e^{-i\gamma}(\hat{\ve{\epsilon}}^*_{3}\cdot\hat{\ve{\epsilon}}_{1})(\hat{\ve{\epsilon}}_{3}\cdot\hat{\ve{\epsilon}}^*_{2})\bra{m}\rho(\ve{q}_2)\ket{n}\bra{ n 
}\rho(-\ve{q}_1)\ket{m}\right] \nonumber \\
& \times \delta(E-E_n+E_m).
\label{eq:ddcsint}
\end{align}
By inspection, we recognize the first two terms as the usual dynamic structure factor,\cite{sinha}
\begin{align}
S(\ve{q},\omega)=\sum_{n,m} b_m |\bra{ n }\rho(\ve{q})\ket{ m }|^2\delta(E-E_n+E_m),
\end{align}
which is the correlation function measured in normal IXS experiments.  Hence, Eq. \ref{eq:ddcsint} can be rewritten 
\begin{align}
\frac{d^2\sigma}{dE' d\Omega}=& \left(\frac{e^2}{mc^2}\right)^2\frac{\omega_3}{\omega_1}\left[g_1^2 
|\hat{\ve{\epsilon}}^*_{3}\cdot\hat{\ve{\epsilon}}_{1}|^2S(\ve{q}_1,\omega)  + g_2^2 |\hat{\ve{\epsilon}}^*_{3}\cdot\hat{\ve{\epsilon}}_{2}|^2 
S(\ve{q}_2,\omega) \right. \nonumber \\
& \left. + g_1 g_2 e^{i\gamma} 
(\hat{\ve{\epsilon}}_{3}\cdot\hat{\ve{\epsilon}}^*_{1})(\hat{\ve{\epsilon}}^*_{3}\cdot\hat{\ve{\epsilon}}_{2})S(\ve{q}_1,\ve{q}_2,\omega) \right. \nonumber \\
& \left. + g_1 g_2 e^{-i\gamma} 
(\hat{\ve{\epsilon}}^*_{3}\cdot\hat{\ve{\epsilon}}_{1})(\hat{\ve{\epsilon}}_{3}\cdot\hat{\ve{\epsilon}}^*_{2})S(\ve{q}_2,\ve{q}_1,\omega)
\right],\label{eq:ddcs}
\end{align}
where $\omega = \omega_1 - \omega_3$ is the energy transferred to the sample, and we have defined a nondiagonal, generalized dynamic structure factor
\begin{align}
S(\ve{q}_1,\ve{q}_2,\omega) & \equiv \displaystyle\sum_{n,m} b_m \bra{m}\rho(\ve{q}_1)\ket{n}\bra{n}\rho(-\ve{q}_2)\ket{m} \delta(\hbar\omega - E_n + E_m ).
\label{eq:offdiagonalS}
\end{align}
This quantity physically represents the degree to which charge fluctuations with wave vector $\ve{q}_1$ are correlated with those with wave vector $\ve{q}_2$.  In the next section we will see that this generalized correlation function is  related to the desired off-diagonal components of the susceptibility, $\chi(\ve{q}_1,-\ve{q}_2,\omega)$.   

\section{Generalized Fluctuation Dissipation Theorem}

To proceed further we need a fluctuation-dissipation theorem that applies to the generalized function, $S(\ve{q}_1,\ve{q}_2,\omega)$.  The retarded response function, $\chi$, is defined in real space as a quantum mechanical thermal average,\cite{fetter}
\begin{align}
\label{eq:ddx}
\chi(\ve{x}_1,\ve{x}_2,t_1-t_2) & = -\frac{i}{\hbar}\displaystyle\sum_m b_m\bra{m}\left[\hat{\rho}(\ve{x}_1,t_1),\hat{\rho}(\ve{x}_2,t_2)\right]\ket{m}\theta(t_1-t_2),
\end{align}
where $\theta(t)$ is a step function.  In reciprocal space the imaginary part is given by
\begin{align}
\operatorname{Im}&\chi(\ve{p}_1,\ve{p}_2,\omega) \nonumber \\
 = & -\pi\displaystyle\sum_{n,m} b_m 
\left[\bra{m}\rho(\ve{p}_1)\ket{n}\bra{n}\rho(\ve{p}_2)\ket{m}\delta(\hbar\omega-E_n+E_m) \right. \nonumber \\
& \left. -\bra{m}\rho(\ve{p}_2)\ket{n}\bra{n}\rho(\ve{p}_1)\ket{m}\delta(\hbar\omega+E_n-E_m)\right]\label{eq:imchik}.
\end{align}
where the $\ket{m}$ are, again, many-body eigenstates of 
the electronic system.  Eqs.~\ref{eq:imchik} and \ref{eq:offdiagonalS} together imply that
\begin{align}
\operatorname{Im}&\chi(\ve{q}_1,\ve{q}_2,\omega) \nonumber \\
 = & -\pi S(\ve{q}_1,-\ve{q}_2,\omega) \nonumber \\
& + \pi\displaystyle\sum_{n,m}b_m\bra{m}\rho(\ve{q}_2)\ket{n}\bra{n}\rho(\ve{q}_1)\ket{m}\delta(-\hbar\omega - E_n + E_m).
\end{align}
Switching the indices $n$ and $m$ in the second term and noting that $b_n = b_m e^{-\beta\hbar\omega}$, we arrive at 
\begin{equation}
S(\ve{q}_1,\ve{q}_2,\omega) = -\frac{1}{\pi}\frac{1}{1-e^{-\beta\hbar\omega}}\operatorname{Im}\chi(\ve{q}_1,-\ve{q}_2,\omega).
\label{eq:genfd}
\end{equation}
This expression is a generalized form of the fluctuation-dissipation theorem that applies to a system that lacks translational symmetry.  This result is significant because it implies that the cross section Eq. \ref{eq:ddcs}, which contains terms involving the generalized $S$, provides access to the needed off-diagonal terms of $\chi$.  Of course, these terms are tangled up with the usual diagonal response, but they may be isolated by varying the phase shift, $\gamma$, subtracting the diagonal structure factor terms, and properly normalizing.\citep{schulke1,schulke2} 

While the general form of  Eq. \ref{eq:genfd} is not unexpected it reveals a crucially important detail, which is that there is a sign difference on the second momentum, $\ve{q}_2$, between $\chi$ and $S$.  As we will see below, this sign is crucial for the viability of standing wave IXS as a technique.

\section{Quantum particle in a periodic potential}

We now address the question of whether standing wave IXS can provide, at least in principle, enough information to carry out a refinement of the full real space response, $\chi(\ve{x}_1,\ve{x}_2,t)$.  We begin by pointing out that, because of the negative sign on $\ve{q}_2$ in Eq. \ref{eq:genfd}, the experimental standing wave constraint, $\ve{k}_1-\ve{k}_2=\ve{G}$, is not as severe as it may seem.  Because the system is periodic, there is also a constraint on the response itself, i.e., 
\begin{equation}
\chi(\ve{x}_1,\ve{x}_2,t)=\chi(\ve{x}_1+\ve{R},\ve{x}_2+\ve{R},t),
\end{equation}
where $\ve{R}$ is a real space Bravais lattice vector.  In reciprocal space this constraint has the form\cite{abbamonte:054302}
\begin{equation}
\chi(\ve{p}_1,\ve{p}_2,\omega)=\chi(\ve{p}_1,\ve{G}-\ve{p}_1,\omega)
\label{eq:chiper}
\end{equation}
i.e., $\chi$ is nonzero only when $\ve{p}_1+\ve{p}_2=\ve{G}$.  In terms of the actual momenta in the experiment, related by Eq. \ref{eq:genfd}, this constraint implies that $\ve{G}=\ve{p}_1+\ve{p}_2$$=\ve{q}_1-\ve{q}_2$$=\ve{k}_1-\ve{k}_3-\ve{k}_2+\ve{k}_3$ or, in other words, that $\chi$ is nonzero only when
\begin{equation}
\ve{k}_1-\ve{k}_2=\ve{G}.
\end{equation}
This is, however, precisely the same as the constraint on the momenta from the standing wave condition itself.

In other words, the minus sign on $\ve{q}_2$ in Eq. \ref{eq:genfd} has the consequence that the standing wave constraint {\it is no constraint at all}.  While the standing wave approach does not allow access to all possible combinations of momenta, $\ve{q}_1$ and $\ve{q}_2$, not all combinations are needed; the response function, $\chi$ is nonzero only for a select subset of momenta, and this subset is exactly what is accessible in the experiment.  

The last issue is whether, at a fixed value of $\ve{G}$, there is sufficient flexibility in the remaining momentum ($\ve{p}_1$ in Eq. \ref{eq:chiper}) to perform a refinement of the complete $\chi$.  To 
answer this question requires a specific, microscopic model.  For this purpose, we consider the case of a single, spinless quantum particle traveling in a periodic array of harmonic wells, 
\begin{equation}
V(\ve{x}) = -\nu r_0^2 + \displaystyle\sum_{\ve{R}}\nu\left|\ve{x}-\ve{R}\right|^2\theta(r_0-\left|\ve{x}-\ve{R}\right|),\label{eq:gaussianv}
\end{equation}
where $r_0$ is the radius and $\nu$ the depth of each well. For simplicity we assume the system is simple cubic with lattice parameter $a$, and that $r_0<a$.  While this model is not particularly physical, the conclusions we draw from it about momentum constraints will be quite general.  

If the wells are deep, the lowest energy band can be computed in the LCAO or ``tight binding" approximation.  Its dispersion is
\begin{equation}
\label{eq:dispersion}
\omega(\ve{k})=\omega_0-2 t_h \left [\cos(k_x a)+\cos(k_y a)+\cos(k_z a) \right ] - \nu r_0^2
\end{equation}
where $\omega_0$ is the ground state energy of an isolated well	 and $-t_h$ is the hopping parameter between wells.  The corresponding bloch waves have the form $\psi_{\ve{k}}(\ve{x})= \sum_{\ve{R}}\phi(\ve{x}-\ve{R})e^{i\ve{k}\cdot\ve{R}}/\sqrt{N}$, where the tight-bonding orbitals are just the ground state of the simple harmonic oscillator, 
\begin{equation}
\phi(\ve{x}) = (\pi\sigma^2)^{-\frac{d}{4}}e^{-\frac{\ve{x}^2}{2\sigma^2}}\label{eq:gaussianorb}
\end{equation}
and $N$ is the number of wells.  The electron field operator is then given by 
\begin{equation}
\hat{\psi}(\ve{x})=\displaystyle\sum_{\ve{k}} c_{\ve{k}} \psi_{\ve{k}}(\ve{x})
\end{equation}
where $c_{\ve{k}}$ annihilates an electron in state $\psi_{\ve{k}}$, and $d$ is the dimensionality ($d$=1, 2, or 3).

We now wish to compute the generalized dynamic structure factor, $S(\ve{q}_1,\ve{q}_2,\omega)$, for this model, and determine$-$for any given value of $\ve{G}$$-$whether it is experimentally possible to access a sufficient range of momenta.  Because the bands disperse, the scattered 
intensity depends not just on the momenta, but also on the value of the transferred energy, $\omega$, resulting in a problem that is effectively seven dimensional.  To simplify the discussion, we define an auxiliary quantity 
\begin{align}
\label{eq:matrix}
\xi(\ve{q}_1,\ve{q}_2) &\equiv \hbar \displaystyle\int^{\infty}_{-\infty} 
d\omega \; S(\ve{q}_1,\ve{q}_2,\omega)\nonumber\\
& = \displaystyle
\sum_{n,m}b_m\bra{m}\hat{\rho}(\ve{q}_1)\ket{n}\bra{n}\hat{\rho}(-\ve{q}_2)\ket{m},
\end{align}
which describes the total amount of spectral weight to be found at a given combination of $\ve{q}_1$ and $\ve{q}_2$, irrespective of the 
value of $\omega$.  For simplicity, we take the zero temperature limit, $b_0=1$ and $b_{m\ne0}$=0, in which case 
\begin{equation}
\xi(\ve{q}_1,\ve{q}_2)=
\displaystyle\sum_{\ve{k}} \bra{0}\hat{\rho}(\ve{q}_1)\ket{\ve{k}}\bra{\ve{k}}\hat{\rho}
(-\ve{q}_2)\ket{0},
\label{eq:matrixfix}
\end{equation}
where we have now labeled the states in terms of the momentum, i.e., $\ket{\ve{k}}$ denotes the electron in eigenstate $\psi_{\ve{k}}(\ve{x})$, $\ket{0}$ being the ground state.

In momentum space, the density operator $\hat{\rho}(\ve{x})=\hat{\psi}^{\dagger}(\ve{x})\hat{\psi}(\ve{x})$ has the form 
\begin{equation}
\hat{\rho}(\ve{q})=\tilde{\Phi}(\ve{q}) \displaystyle\sum_{\ve{k},\ve{G}}c_{\ve{k}}^{\dagger}c_{\ve{k}+\ve{q}+\ve{G}}\label{eq:rhoq}
\end{equation}
where $\tilde\Phi(\ve{q})$ is the Fourier transform of $|\phi(\ve{x})|^2$. 
Substituting Eq.~\ref{eq:rhoq} into Eq.~\ref{eq:matrixfix} gives
\begin{align}
\xi(\ve{q}_1,\ve{q}_2) & = N \tilde\Phi(\ve{q}_1)\tilde\Phi(\ve{q}_2)\displaystyle\sum_{\ve{G}}\delta_{\ve{q}_1-\ve{q}_2,\ve{G}}
\label{eq:xifinal}
\end{align}
where we have used the fact that $|\phi(\ve{x})|^2$ is centrosymmetric, i.e., $\tilde\Phi(\ve{q}) = \tilde\Phi(\ve{-q})$.  As expected, the intensity scales with the number of scatterers, $N$, and is nonzero only when $\ve{q}_1-\ve{q}_2=\ve{G}$, which required by periodicity.  Subject to this constraint, the intensity is just given by the quantity $\tilde\Phi(\ve{q}_1)\tilde\Phi(\ve{q}_2)$.  We now evaluate where this product is nonzero and whether it can be adequately sampled in a physical experiment.


\subsection{One-Dimensional Case}

For illustration, it is useful to consider the problem first in one dimension, before moving on to the physical cases of two and then three dimensions.  

In one-dimension Eq.~\ref{eq:gaussianv} describes a chain of harmonic wells.  This case is not experimentally realizable, since it requires the two momentum transfers to be collinear, which is impossible except in the limit of infinite energy.  We can, however, draw several important conclusions about the functional form of $\xi$ from this case.  

\begin{figure}\centering
\subfigure[][]{\includegraphics[width=0.4\textwidth]{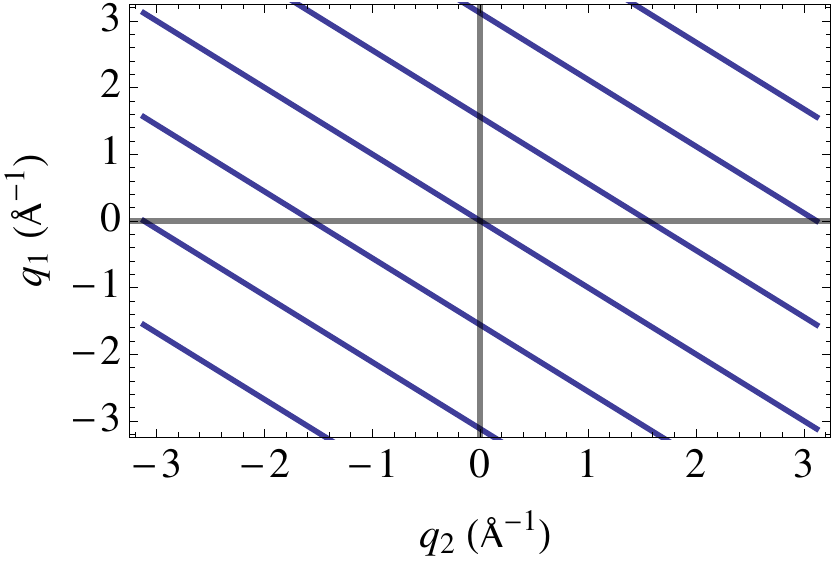}
\label{fig:diag}}
\subfigure[][]{\includegraphics[width=0.5\textwidth]{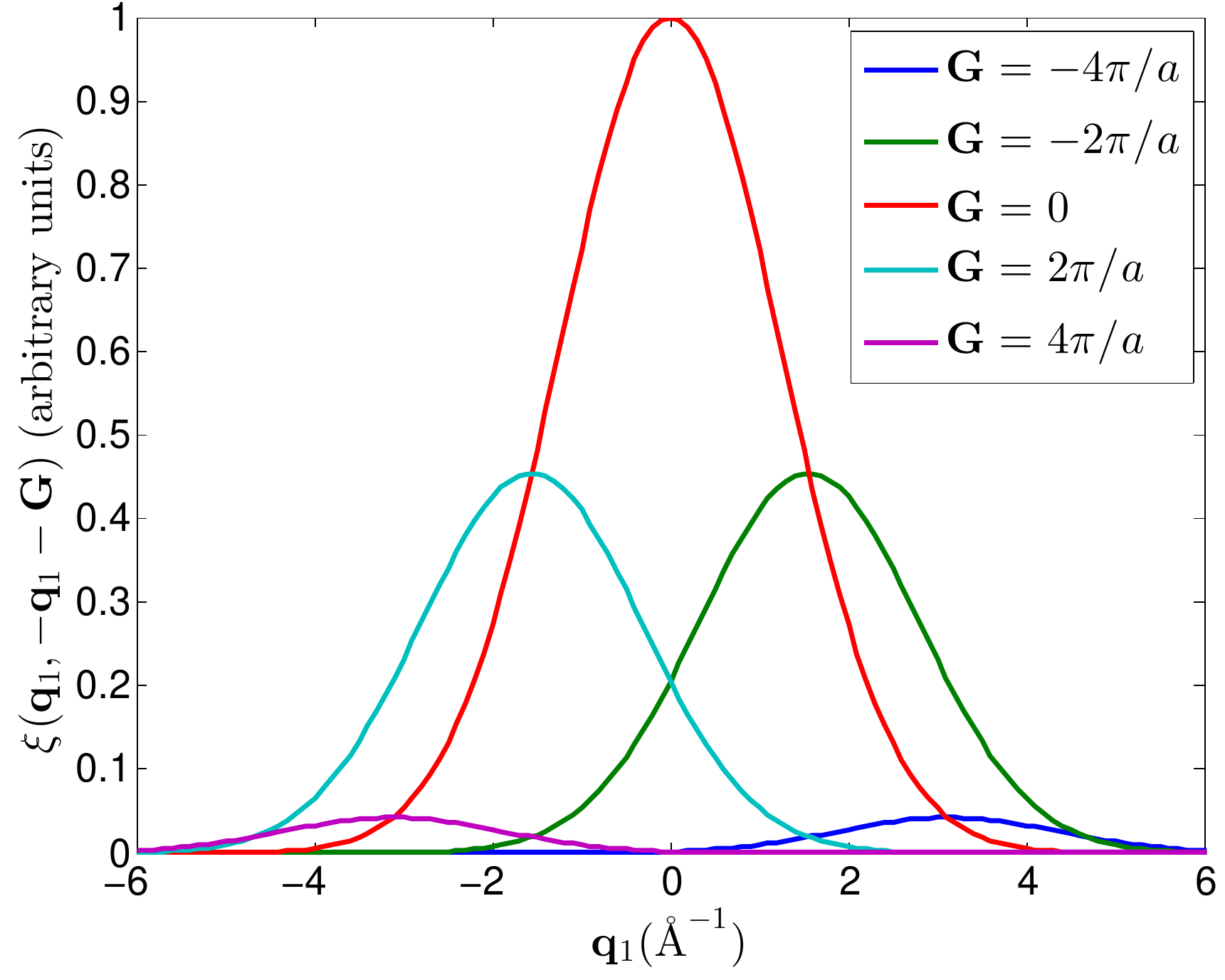}
\label{fig:gauss}}%
\caption{\subref{fig:diag} Lines in the $(q_1,q_2)$ plane along which $\xi(q_1,q_2)$ may be non-zero in the one-dimensional case.  \subref{fig:gauss} Value 
of $\xi(q_1,q_1-G)$ for a few selected values of $G$.}
\label{fig:oned}
\end{figure}

In one dimension the momenta $q_1$ and $q_2$ are scalars, and the reciprocal lattice vectors $G=2\pi h/a$ are indexed by a single integer, $h$.  The resulting value of $\xi(q_1,q_2)$ is illustrated in Fig.~\ref{fig:oned}.  Because of the constraint $q_1-q_2=G$, $\xi$ is nonzero only along the set of lines depicted in Fig.~\ref{fig:diag}. 
 
Fig.~\ref{fig:gauss} shows the magnitude of $\xi(q_1,q_1-G)$ plotted against $q_1$ along a selection of these lines.  Two important 
observations can be made from this plot.  First, the intensity along a given line is not uniform, but has the shape of a Gaussian centered half-way between 
$q_1=0$ and $q_1=G$, whose momentum width $\Delta q_1=\sigma^{-1}$.  Second, the intensity varies from one line to the next, decaying with increasing $|G|$, again with a Gaussian envelope with width $\sigma^{-1}$.  

From this simple case we see an important relationship between the degree of inhomogeneity of the system and the range of momenta that must be sampled 
experimentally to reconstruct $\chi(x_1,x_2,t)$.  The narrower the wells, i.e., the more inhomogeneous the system, the larger the number of $G$ values that 
must be sampled, and the larger the range of $q_1$ that must be measured for each $G$.  Hence, the number of $G$ values that must be sampled is not infinite, but is of order $a/\sigma$, which is a measure of the strength of local field effects.  For the parameters chosen here, one need only sample up to $h=2$ to acquire $> 99 \%$ of the spectral weight that is available.  In the limiting case of a homogeneous system, $\sigma \rightarrow \infty$, $\xi$ is nonzero only for $G=0$, and conventional IXS can completely parameterize the response, as expected.

\subsection{Two-Dimensional Case}

\begin{figure}\centering
\begin{tikzpicture}[scale=1]
\path (0,0) coordinate (origin);
\path (180:3.12cm) coordinate (G020);
\path (90:3.12cm) coordinate (G200);
\path (180-47.78/2:3.85cm) coordinate (kB);
\path (180+47.78/2:3.85cm) coordinate (kI);
\path (180+47.78/2-10:3.85cm) coordinate (kD);
\path (0.217cm, -0.635cm) coordinate (qD);
\path (0.22cm, 2.48cm) coordinate (qB);
\draw[dashed, ->] (origin) -- (G020) node[left=2mm]{$\frac{2\pi}{a}(0,2)$};
\draw[dashed, ->] (origin) -- (G200) node[above=2mm]{$\boldsymbol{G}=\frac{2\pi}{a}(2,0)$};
\draw[thick, ->] (origin) -- node[left=2cm,above=1cm]{$k_2$}(kB);
\draw[thick, ->] (origin) -- node[left=2cm,below=1cm]{$k_1$}(kI);
\draw[thick, ->] (origin) -- node[left=2.5cm,below=2mm]{$k_3$}(kD);
\draw[thick, blue](180:2cm) arc (180:180+47.78/2:2cm) node[left=2mm]{$\eta$};
\draw[thick, red] (180+47.78/2-10:3cm) arc (180+47.78/2-10:180+47.78/2:3cm) node[left=1mm]{$\varphi$};
\draw[very thick, green,->] (origin) -- node [right=2mm]{$q_1$} (qD);
\draw[very thick, green,->] (origin) -- node [right=5mm,above=1cm]{$q_2$} (qB);
\end{tikzpicture}
\caption{Relative magnitude and orientation for the experimental geometry in the two-dimensional, co-planar case.  Here $\ve{G}=2\pi(2,0)/a$.}
\label{fig:vectors}
\end{figure}
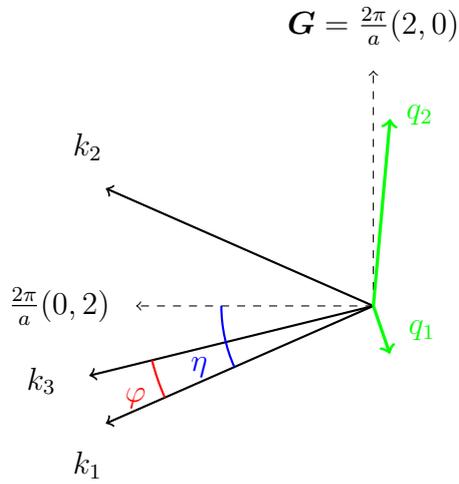

In two dimensions, a standing wave experiment is physically realizable, and corresponds to all the rays in Fig.~\ref{fig:genexp} lying in a single plane.  In this case Eq.~\ref{eq:gaussianv} describes a square, planar array of harmonic wells, the quantities $\ve{q}_1$ and $\ve{q}_2$ are 
two-component vectors, and the reciprocal lattice vectors $\ve{G}=2\pi(h,k)/a$ are described by two indices, $h$ and $k$.  

\begin{figure}\centering
\subfigure[$\ve{G}=\frac{2\pi}{a}(0,2)$]{\includegraphics[scale=0.5]{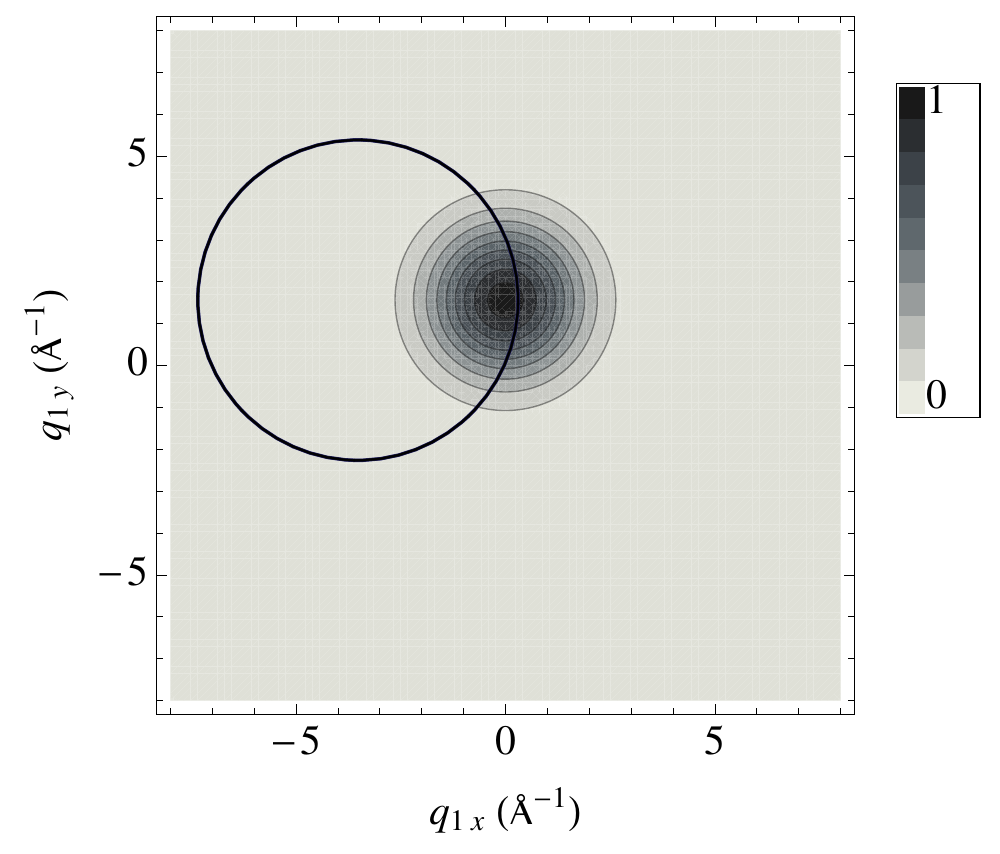}\label{fig:cut020}}
\subfigure[$\ve{G}=\frac{2\pi}{a}(2,0)$.]{\includegraphics[scale=0.5]{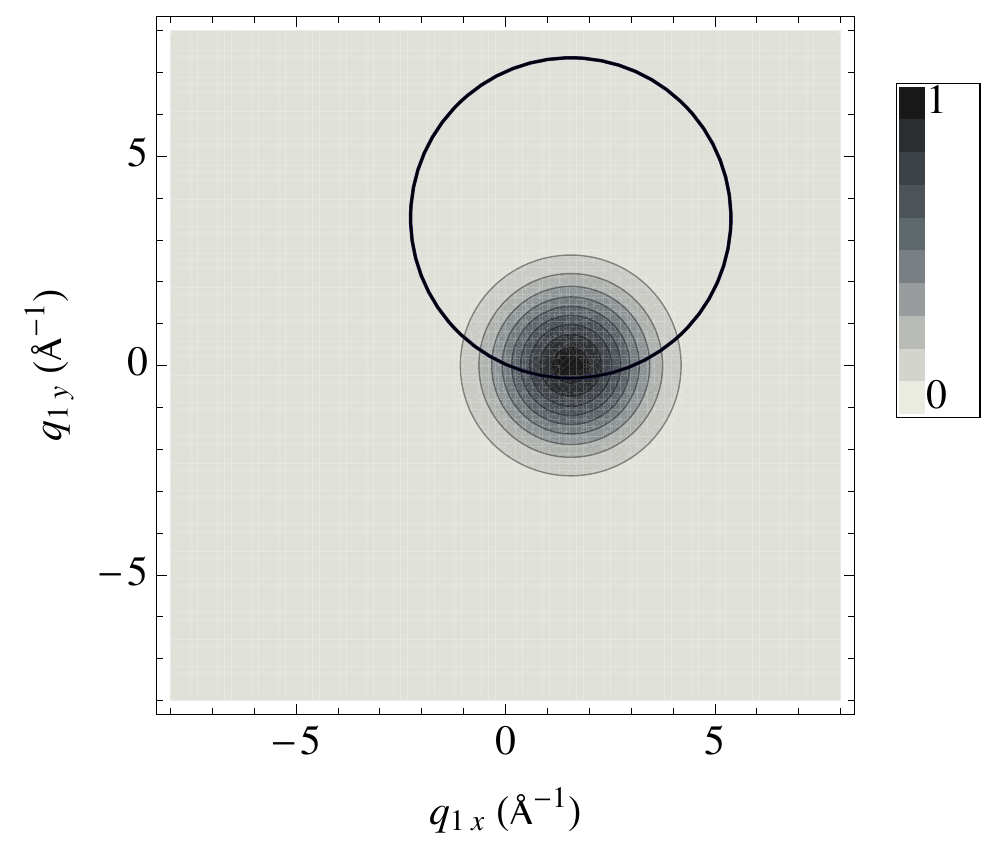}\label{fig:cut200}}
\subfigure[$\ve{G}=\frac{2\pi}{a}(-2,0)$]{\includegraphics[scale=0.5]{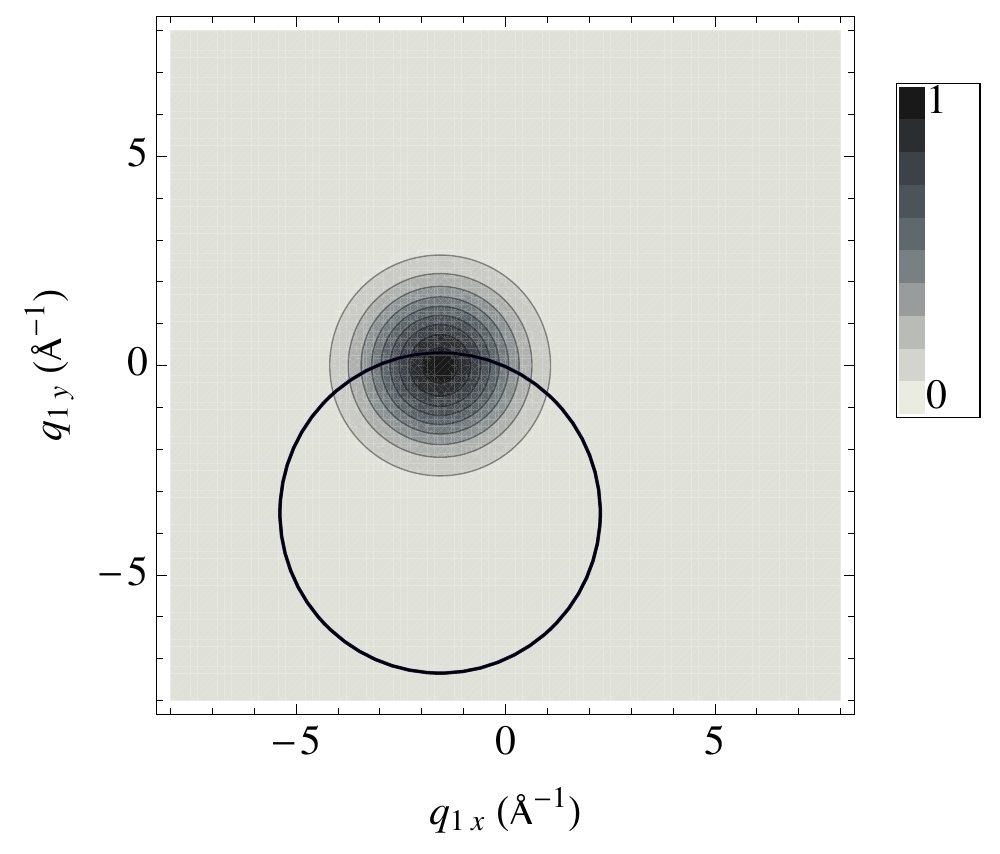}\label{fig:cut_200}}
\subfigure[$\ve{G}=\frac{2\pi}{a}(0,-2)$]{\includegraphics[scale=0.5]{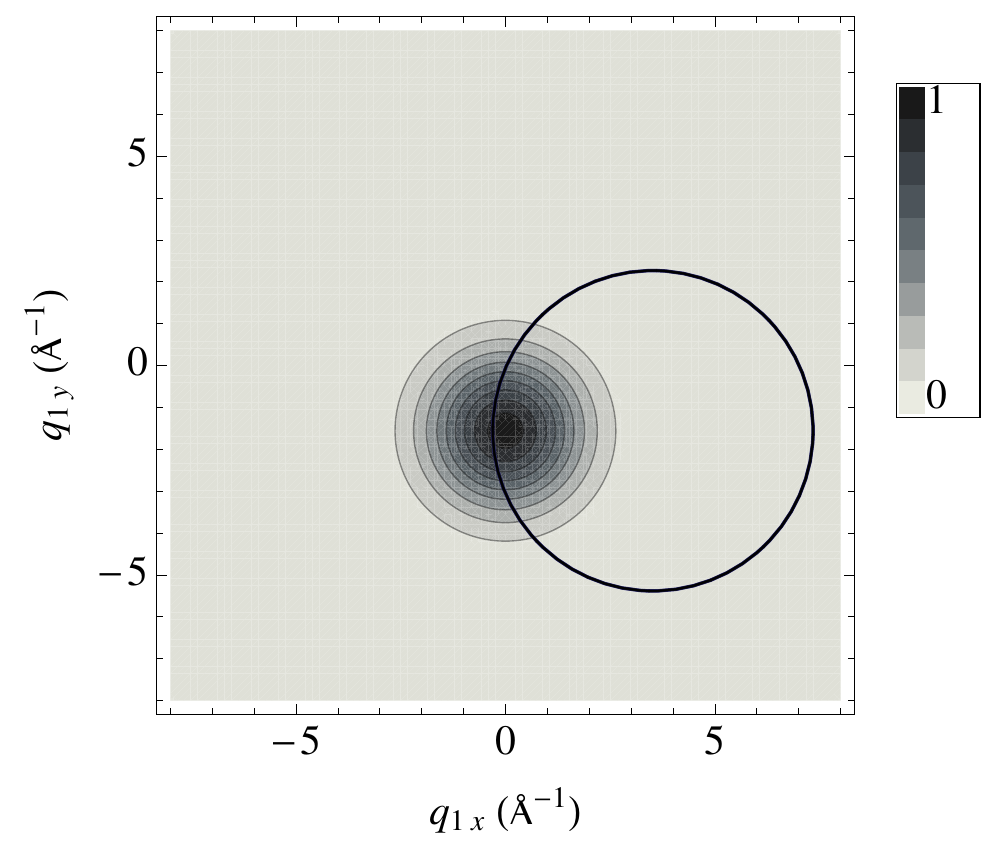}\label{fig:cut0_20}}
\caption{$\xi(\ve{q}_1,\ve{q}_1-\ve{G})$ plotted for the two-dimensional case against the two components $q_{1x}$ and $q_{1y}$ for several values of 
$\ve{G}$.  $\xi$ has the form of a Gaussian centered at the half-way point between $\ve{q}_1=0$ and $\ve{q}_1=\ve{G}$.  The solid lines are the contours 
traced out by rotating the IXS analyzer through 360$^\circ$.}
\label{fig:cuts}
\end{figure}

The quantity $\xi(\ve{q}_1,\ve{q}_2)$ is now a function in four-dimensional space and is not easily illustrated.  Extending reasoning from the 
one-dimensional case, however, we expect $\xi$ to be nonzero only along discrete, two-dimensional sections through this space, again defined by the 
constraint $\ve{q}_1-\ve{q}_2=\ve{G}$.\footnote{Generally speaking, in $d$ dimensions $\xi$ is a scalar function residing in a $2d$-dimensional space, and the constraint $\ve{q}_1-\ve{q}_2=\ve{G}$ defines a discrete set of $d$-dimensional sections through this space.}  Moreover, we expect the magnitude of $\xi$ to decrease with increasing $|\ve{G}|$, and to be 
substantial only for $\sqrt{h^2+k^2} \leq 2$.  By analogy with Fig.~\ref{fig:gauss}, we plot in Fig.~\ref{fig:cuts} several sections of 
$\xi(\ve{q}_1,\ve{q}_1-\ve{G})$ for selected values of $\ve{G}$.  As in Fig.~\ref{fig:gauss}, $\xi$ is a Gaussian function centered at the midpoint between 
$\ve{q}_1=0$ and $\ve{q}_1=\ve{G}$, whose width is isotropic and equal to $\sigma^{-1}$.  

Because the two-dimensional case is experimentally realizable, it is now possible to address the fundamental question (see Section \ref{sec:intro}) of 
whether enough information is accessible with standing wave IXS to reconstruct $\chi(\ve{x}_1,\ve{x}_2,t)$.  To do so, one must experimentally parameterize a 
set of two-dimensional surfaces, such as those illustrated in Fig.~\ref{fig:cuts}, each defined by a distinct $\ve{G}$.  


For a given choice of $\ve{G}$, the angle of the beam with respect to the crystal is fixed by the Bragg condition.  Hence, the only means of adjusting the 
momenta is to adjust the scattering angle of the IXS analyzer.  Doing so has the effect of tracing out a 
circular trajectory in momentum space, that has radius $k_1$ and intersects both the origin and $\ve{q}_1=\ve{G}$.  For illustrative purposes we 
display in Fig.~\ref{fig:cuts} (black lines) each of these trajectories, for the specific case of an x-ray beam energy $E= 15.5 \, \hbar c/a$.

A serious problem is now evident.  One must experimentally parameterize two-dimensional surfaces such as those shown in Fig.~\ref{fig:cuts}, however the set of experimentally accessible points is only one-dimensional, corresponding to a cut through each of the needed surfaces. Hence, we are faced with a crisis of dimensionality: The set of accessible information is of lower dimensionality than what is needed to refine the response function.

\begin{figure}\centering
\includegraphics[scale=0.5]{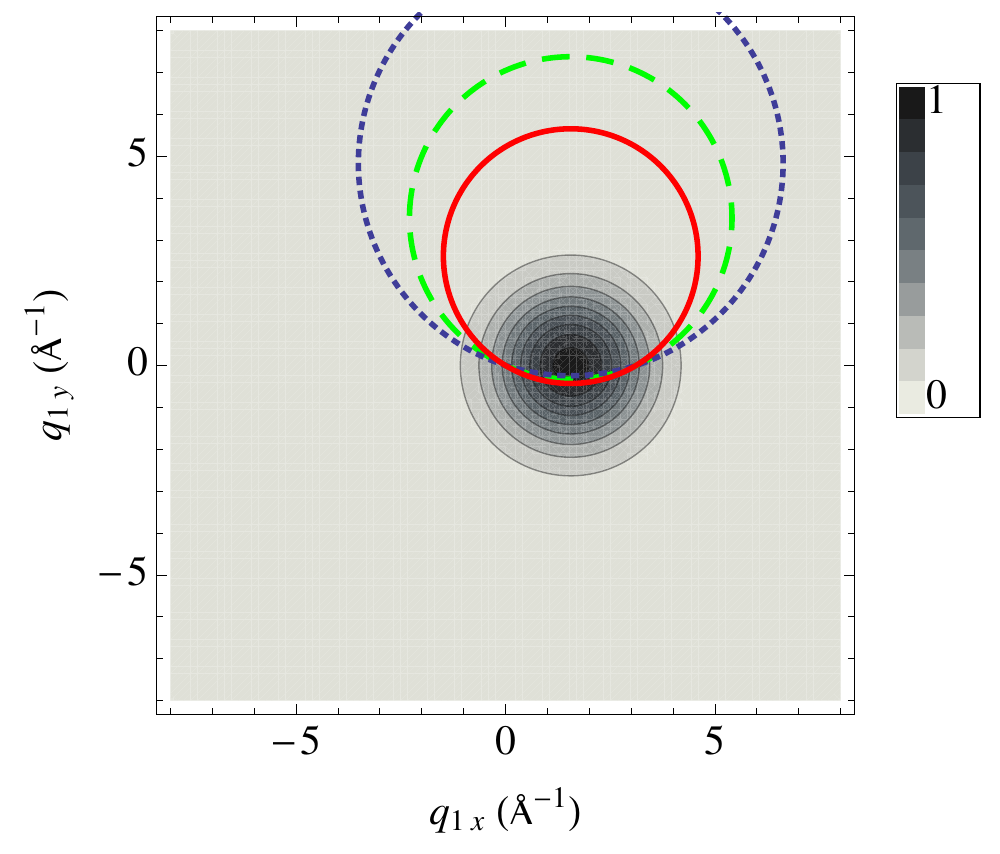}
\caption{Contour shown in Fig. 5(b) for several values of the beam energy.  Red, dashed green, and dotted blue contours correspond to $E=12.3\,\hbar c/a$, 
$E=15.5\,\hbar c/a$, and $E=20.4\,\hbar c/a$, respectively.  Continuously tuning the beam energy in this manner can, in principle, trace out a two-dimensional surface, though doing so is prohibitively difficult experimentally.}
\label{fig:multicut}
\end{figure}

The dimensionality could be increased, of course, by adjusting also the beam energy, as illustrated in Fig.~\ref{fig:multicut}.  This changes the radius of 
the circular section, in principle allowing one to sweep out a two-dimensional surface.  Unfortunately, the nature of IXS is such that each incident energy 
(apart from modest amounts of tunability) requires a distinct experimental setup with a distinct energy analyzer.  We conclude that the two-dimensional case 
of standing wave IXS imaging, while experimentally realizable, lacks the momentum flexibility to be viable for imaging.  

\subsection{Three-Dimensional Case}

Finally, we address the three-dimensional case, in which Eq.~\ref{eq:gaussianv} describes a cubic lattice of harmonic wells.  All momenta now have three 
components, and the $\ve{G}$ vectors are described by three integers, $h$, $k$, and $l$.  $\xi(\ve{q}_1,\ve{q}_2)$ is a function in six-dimensional 
space, but periodicity dictates that it is nonzero only along discrete, three-dimensional contours parameterized by $\ve{q}_1-\ve{q}_2=\ve{G}$.  By analogy 
with the previous two cases, we expect the function $\xi(\ve{q}_1,\ve{q}_1-\ve{G})$ to be a Gaussian with width $\sigma^{-1}$ centered on the point 
$\ve{q}_1=\ve{G}/2$.

In terms of the measurement itself, two types of motions are now possible that were not available in either of the previous cases: (1) The analyzer may now 
be rotated in two directions, i.e., both parallel and perpendicular to the Bragg plane. (2) The sample may be rotated around an axis parallel to $\ve{G}$, changing the sample angles while maintaining the Bragg condition.  The latter motion is usually referred to as a ``$\Psi$ rotation".

The analyzer motion, which is now two-dimensional, traces out a spherical shell in $\ve{q}_1$ space.  As in the two-dimensional case, this sphere has radius 
$k_1$ and intersects both the origin and the point $\ve{q}_1=\ve{G}$.  As before, this section is of lower dimensionality than the space spanned by the 
function $\xi(\ve{q}_1,\ve{q}_1-\ve{G})$.  An additional degree of freedom is needed to make the dimensionality of the measurement match the that of the 
$\xi$ function.

This degree of freedom is supplied by the $\Psi$ rotation.  Rotating the crystal around the Bragg vector (i.e., maintaining the Bragg condition) has the 
effect of sweeping the spherical shell around an axis defined by the line connecting the origin and $\ve{q}_1=\ve{G}$.  In this manner, the shell sweeps out 
a {\it three-dimensional volume} in momentum space.  The volume swept out is a torus, with major radius $\sqrt{{k_1}^2-G^2}$ and minor radius $k_1$, as 
illustrated in Fig. 7.  A plot of the intersection of this torus with a constant-contour plot of $\xi(\ve{q}_1,\ve{q}_1-\ve{G})$ is shown in 
Fig.~\ref{fig:3dstuff}.  We conclude that the crisis of dimensionality encountered in the two-dimensional case does not take place here, and that standing 
wave IXS imaging$-$in principle$-$should be a viable technique in the real world of three dimensions.

While the dimensionality of the experiment is adequate, the three dimensional case has some blind spots.  As one might expect, information outside the torus 
shown in Fig.~\ref{fig:3dstuff} is not accessible.  This is the standing-wave manifestation of the well-known diffraction limit, which says that it is not 
possible in a scattering experiment to extract information about features smaller than half a wavelength.  

\begin{figure}\centering
\includegraphics[scale=0.5]{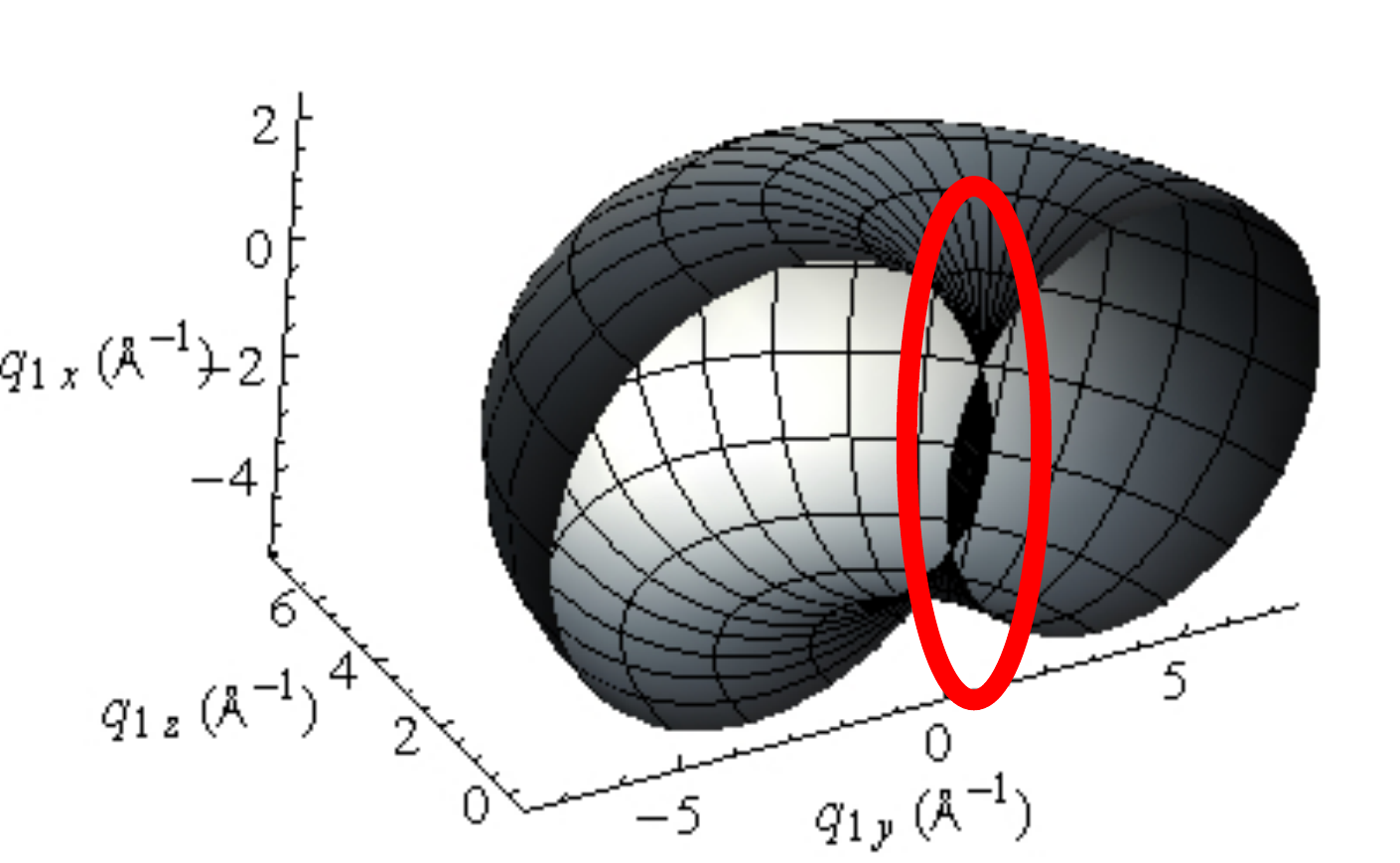}
\caption{Toroidal volume swept out by combined rotation of the two detector motions and the sample azimuth, $\Psi$, in the three-dimensional case.  The center contains a blind spot whose shape is a {\it vesica piscis} swept through its major symmetry axis.  This blind spot may be made arbitrarily small by increasing the beam energy.}
\label{fig:toroidhalf}
\end{figure}

\begin{figure}\centering
\subfigure[][]{\includegraphics[scale=0.5]{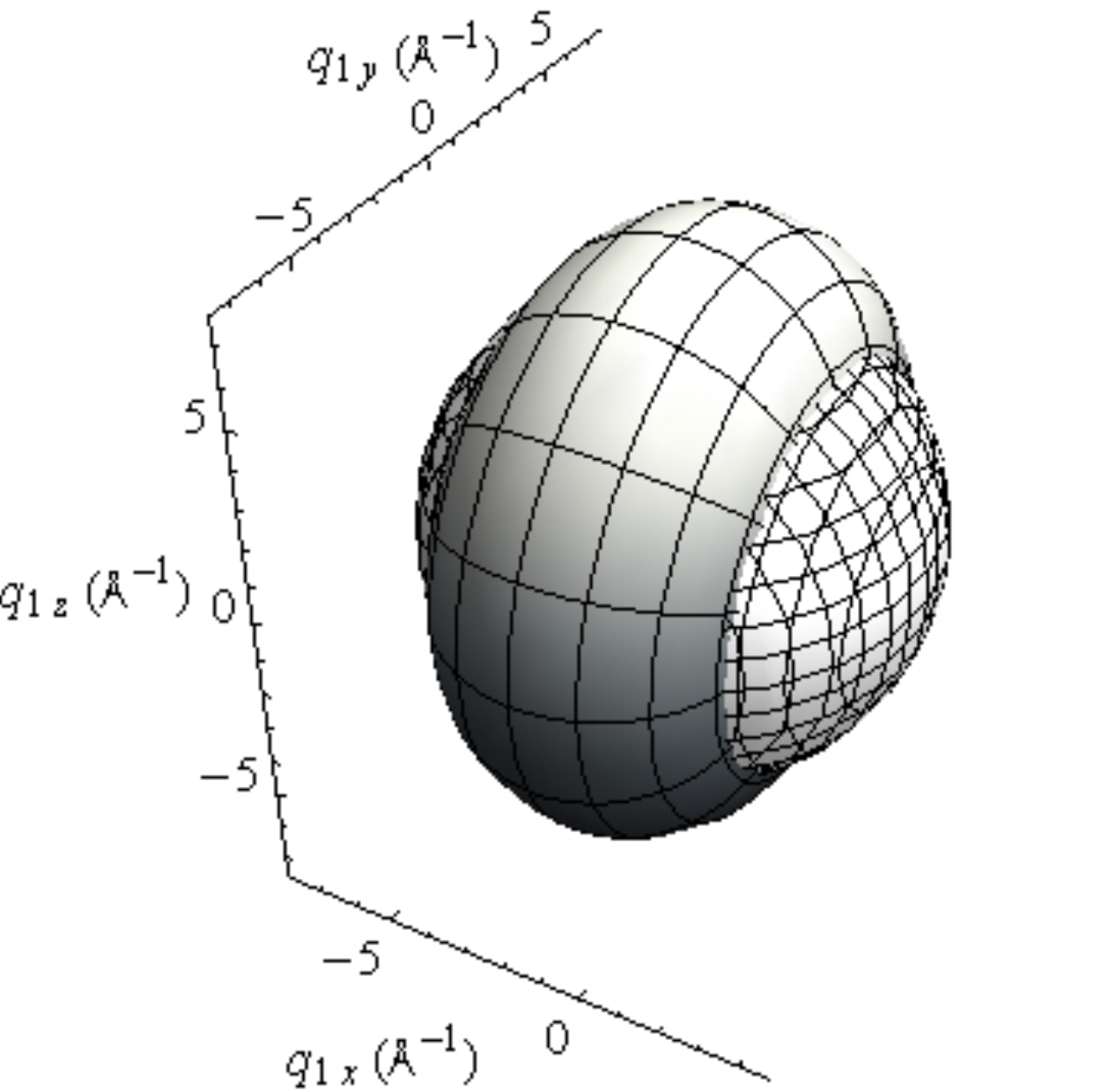}\label{fig:azimuthcut}}
\caption{Intersection of the toroidal measurement volume shown in Fig. 7 with a constant-contour plot of $\xi(\ve{q_1},\ve{q_1}-\ve{G})$.}
\label{fig:3dstuff}
\end{figure}

Second, and less intuitively, there is a blind spot in the center of the torus (see Fig.~\ref{fig:toroidhalf}) whose shape is a three-dimensional 
{\it vesica piscis} with axis length $|\ve{G}|$ and major radius $k_1-\sqrt{{k_1}^2-G^2}$.  It remains to be determined whether this inner blind spot poses a serious limitation on using standing wave IXS for imaging.  We note, however, that the radius of the blind spot goes to zero as $k_1 \rightarrow \infty$, so it can always be made arbitrarily small experimentally by working at sufficiently high beam energy.  

\section{Conclusions}
\label{sec:conclusions}

To summarize, we have found that standing wave measurements are, in principle, a viable approach to overcoming the translational averaging 
problem\citep{abbamonte:054302} in IXS imaging.\citep{water,advmat}  If successfully implemented, this approach would reveal the complete density response, 
$\chi(\ve{x}_1,\ve{x}_2,t)$, which describes the electron disturbance created by a source placed at any arbitrary location, $\ve{x}_1$, in a spatially 
inhomogeneous but periodic system$-$typically with Angstrom spatial and attosecond time resolution.  This technique can be thought of as a generalization of x-ray crystallography that allows refinement of the {\it excited states} of a periodic system, rather than just its ground state, and represents the maximum that can be learned by interaction of light with matter in the regime of linear response.  This technique would be most useful for imaging excitations in very inhomogeneous systems, such as molecular crystals, in which local field effects are significant and transverse collective excitations, such as transverse plasmons, can be important.  

Analyzing a simple model of a single quantum particle in a periodic potential, we have shown that standing wave IXS imaging is an innately three-dimensional 
measurement, in the sense that both out-of-plane analyzer motions and sample $\Psi$-rotations are required to achieve a complete data set.  Such an 
experiment would require (at the minimum) two copies of the crystal of interest: one that is asymmetrically cut to collimate the beam, and a second to create 
the standing wave field and function as the ``sample".  The scattering experiment would then require six rotation axes: a conventional three-axis sample 
goniometer, a single-axis, two-theta rotation supporting an energy-integrating detector for measuring the Bragg diffracted beam, and a second two-theta 
rotation, with both in- and out-of-plane degrees of freedom, supporting a backscattering, IXS analyzer.

One might think that the data collection time required to refine, for example, the three-dimensional $\chi(\ve{x}_1,\ve{x}_2,t)$ for a 
molecular crystal would be enormous, but this is not necessarily so.  Specifically, suppose one were studying a molecular crystal whose ground state density had been refined with 
conventional x-ray crystallography, to a resolution providing $N$ cartesian voxels in each unit cell.  Such a refinement would have required of order $N$ 
independent Bragg measurements (unless other information about the structure, such as its symmetry, were known).  To refine $\chi(\ve{x}_1,\ve{x}_2,t)$ for 
this structure with the same resolution, because the response is a function of both source and observation coordinates, would require of order $N^2$ 
measurements {\it at a single time slice}.  To construct the complete dynamics, then, requires $N^2 N_t$ measurements, where $N_t$ is the number of points in 
the time series.  This number is, without a doubt, impracticably large in nearly all conceivable cases.  

The time considerations, however, need not be so severe.  For example, if one fixes $\ve{x}_1$, i.e., if one decides before the measurement where to 
``strike" the molecule, the number of unknowns reduces to $N N_t$.  Further, as in x-ray crystallography, knowledge of symmetry further reduces the number of 
unknowns.  For example, all four of the independent measurements in Fig.~\ref{fig:multicut}, which are related by symmetry, would reveal the same 
experimental result; this could have been anticipated ahead of time from the crystal symmetry.  Other tricks, such as taking one- or two-dimensional 
projections, compromising the resolution by probing only small order $\ve{G}$ values, etc., are always possible in specific cases.  Hence, we expect the data 
collection time to be no more intensive than the spatially-averaged IXS imaging, which has already been demonstrated.\citep{water,LiF,reed}

\section{Acknowledgements}

We gratefully acknowledge Wei Ku, Thomas Gog, and Diego Casa for helpful discussions.  This work was supported by U.S. Department of Energy grant DE-FG02-07ER46453 through the Frederick Seitz Materials Research Laboratory.  Fig.~\ref{fig:graphite} was generated from data taken at Sector 9 at the Advanced Photon Source, which is supported by DOE grant DE-AC02-06CH11357.

\bibliography{off-diagonal7}{}
\bibliographystyle{elsarticle-num-names}
\end{document}